\documentclass[pre]{revtex4}
\usepackage[dvips]{graphics}

\let\gam\gamma
\let\rh\rho
\let\p\partial
\let\~\tilde
\begin{document}
\title{Granular clustering in a hydrodynamic simulation}
\author{Scott A. Hill}
\author{Gene F. Mazenko}
\affiliation{James Franck Institute and Department of Physics, University of Chicago, Chicago, Illinois 60637.}
\date{\today}
\begin{abstract}
We examine the hydrodynamics of a granular gas using numerical
simulation.  We demonstrate the appearance of shearing and clustering
instabilities predicted by linear stability analysis, and show that their
appearance is directly related to the inelasticity of collisions in
the material.  We discuss the rate at which these instabilities arise
and the manner in which clusters grow and merge.
\end{abstract}
\maketitle

One of the key differences between a granular material and a regular
fluid is that the grains of the former lose energy with each
collision, while the molecules of the latter do not.  Even when the
inelasticity of the collisions is small, it can give rise to dramatic
effects, such as the \textit{Maxwell Demon} effect\cite{eggers} and,
the topic of this paper, the phenomenon of granular clustering.
Experiments\cite{kudrolli,olafsen} and molecular dynamics
simulations\cite{goldhirsch} alike show that low-density collections
of grains (``granular gases'') in the absence of gravity do not become
homogeneous with time, but instead form denser clusters of stationary
particles surrounded by a lower density region of more energetic
particles.  A kinetic explanation for this behavior is that, when a
particle enters a region of slightly higher density, it has more
collisions, loses more energy, and so is less able to leave that
region. This increases the local density and makes it more likely that
additional particles are captured in the same way.

We are interested in describing this clustering behavior using
hydrodynamics.  There is considerable work\cite{kinetic} deriving
granular hydrodynamics from kinetic theory, focusing on analytical
treatments of the long-wavelength behavior of the system.  Goldhirsch
and Zanetti\cite{goldhirsch}, for instance, describe clustering as the
result of a hydrodynamic instability: a region of slightly higher
density has more collisions, so more energy is lost and the region has
a lower ``temperature''\cite{temperature}.  Less temperature results
in less pressure, and this lower pressure region in turn attracts more
mass from the surrounding higher-pressure regions.  Their paper uses
long-wavelength stability analysis to show that, in a system of
hydrodynamic equations similar to Eq.~\ref{eq-eqns} below,
higher-density regions do indeed have lower pressure, fueling the
instability.

In this paper, we study the hydrodynamics of granular clustering in
zero gravity, by using numerical simulation.  Our motivation is to
determine whether a coarse-grained description, in terms of local
particle, momentum, and energy densities, can be used to treat
characteristic behaviors of granular materials as a self-contained
dynamical system\cite{fn-turbulence}.  We show that the instabilities
predicted by linear analysis do arise in our simulations, and discuss
how the onset of these instabilities depends on the inelasticity of
collisions in the material.  We also show the manner in which clusters
develop.

We begin with a number density field $\rho$, a flow velocity field
$\mathbf{u}$, and a ``temperature''\cite{temperature} field $T$.
These are related by a standard set of hydrodynamic equations for
granular materials, introduced by Haff\cite{haff}:

\begin{eqnarray}
\label{eq-eqns}
{\p\rho\over\p t}&=&-\nabla_i(\rho u_i)\cr
{\p(\rho u_i)\over\p t}&=&-\nabla_i P
-\nabla_j(\rho u_iu_j)+\nabla_j(\eta_{ijkl}\nabla_ku_l)\cr
{\p T\over\p t}&=&-\nabla_i(u_iT)+{1\over\rho}\nabla_i(\~\kappa \nabla_i T)\cr
&&\hskip0.5in+\frac{1}{\rho}\eta_{ijkl}(\nabla_iu_j)(\nabla_ku_l)-\~\gam T.
\end{eqnarray}
where repeated indices are summed over, and where $P$ is the pressure,
$\~\kappa$ is the bare thermal conductivity, and
$\eta_{ijkl}=\~\eta(\delta_{ik}\delta_{jl}+\delta_{il}\delta_{kj}+\delta_{ij}\delta_{kl})$
is the isotropic bare viscosity tensor.  These equations bear much in
common with those for normal fluids\cite{kim}.  The most important
addition is that of the term $-\~\gamma T$, which accounts for the
inelasticity of collisions; the parameter $\~\gamma$ is proportional to
$(1-r^2)$, where $r$ is the coefficient of restitution.  Using kinetic
theory results\cite{haff}, the transport coefficients are chosen to
depend on temperature and density:
\begin{eqnarray}
\~\kappa=\kappa T^{1/2}\cr
\~\eta=\eta \rho T^{1/2}\cr
\~\gamma=\gam T^{1/2}.
\end{eqnarray}

Typically, work in granular hydrodynamics is done in low-density
regimes, where grains may be treated as point particles interacting
via collisions.  When simulating aggregation, however, one must take
excluded volume into account.  We do this by introducing a barrier in
the pressure $P(\rho)$ at some maximum (close-packed) density
$\rho_0$.  This is in addition to the usual hydrodynamic pressure
$\rho T$.  We choose in particular the simple quadratic form
\begin{equation}
P=\rho T+U(\rho^2-\rho_0^2)\theta(\rho-\rho_0),
\end{equation}
where $U$ is a positive parameter, $\theta(x)$ is the unit step
function, and $\rho_0$ is the close-packed density.
This method, which we introduced in an earlier paper\cite{hill}, is a
simple way to model the incompressibility of the system at high
densities\cite{gollub}.  

We evaluate our equations in two dimensions using a finite-difference
Runge-Kutta method, on a square lattice with periodic boundary
conditions.  (See footnote \cite{fn-method} for more details.)  The
lattice spacing is chosen to be large enough so that each site
contains a number of grains, and we can consider the density to be a
continuous variable.  We start with random initial conditions
$\rho=0.1+0.001r_1(z,x)$, $u_z=r_2(z,x)$, $u_x=r_3(z,x)$, and
$T=1+0.1r_4(z,x)$, where $r_i(z,x)$ are random numbers chosen between
$-1$ and $1$.  The model's other parameters for the data presented
here are $\eta_0=25$, $\kappa_0=1$, $U=4\times 10^4$, $\rho_0=0.2$,
and $\gam$ taking on several different values.  All numbers given here
are in dimensionless units\cite{fn-units}. Our time step in these
units is $\Delta t=10^{-3}$.

We begin with a system that is $64\times64$ in size.  The homogeneous
state with which we initialize our system is already a solution to the
equations above.  In this initial homogeneous cooling state, the
velocity and all gradients vanish, and the temperature decays with
time due to the inelasticity.  Eq.\ref{eq-eqns} reduces to
\begin{equation}
\frac{\p T}{\p t}=-\gam T^{3/2},
\end{equation}
which yields {\it Haff's cooling law}, $T(t)=T(0)(1+t/t_0)^{-2}$.
This state is seen universally in simulations\cite{DB97, LH99, NBC02,
McY96}, but only initially, for it is unstable to hydrodynamic
modes\cite{goldhirsch}, resulting in a long-range shear flow followed
by the clustering instability mentioned above.

\begin{figure}
\begin{center}\includegraphics{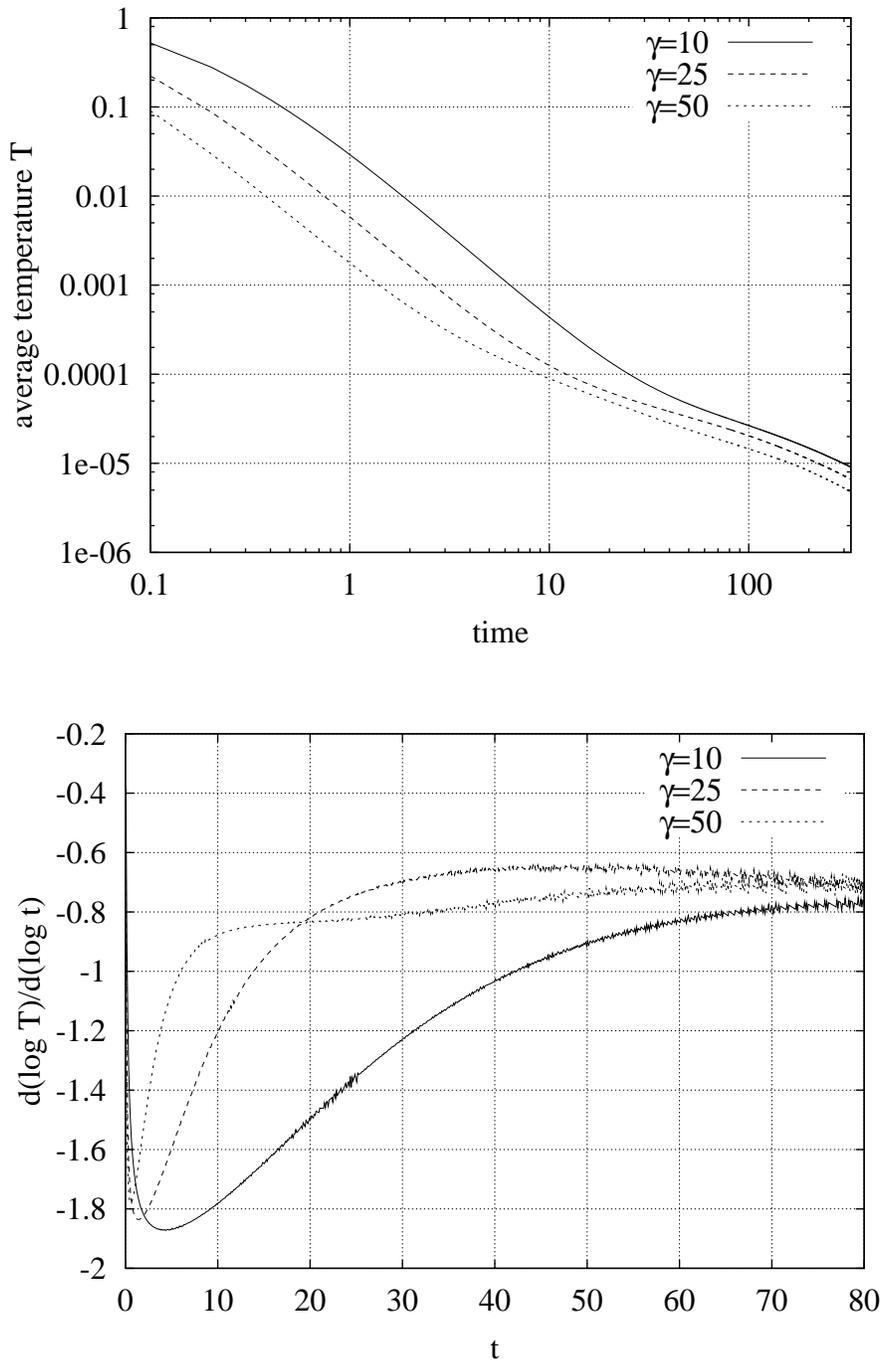}\end{center}
\caption{\label{fig-Tvt}The first plot shows the evolution of the
average temperature of the system over time, for three different
values of the inelasticity parameter $\gam$, on a log-log plot.  The
second is the logarithmic derivative of $T(t)$; that is, $\frac{d(\log
T)}{d(\log t)}$, or the slope of the lines in the first graph.  The
lower graph shows the exponent of the power-law decay rate.}
\end{figure}
\begin{figure}
\begin{center}\includegraphics{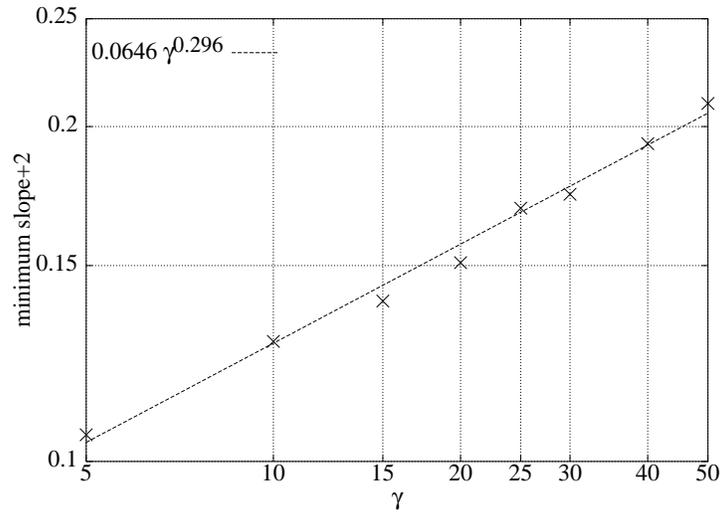}\end{center}
\caption{\label{fig-approachHaff}The maximum power-law decay rate of
the average temperature (i.e. the minimum slope in a $\log T$
vs. $\log t$ plot such as in Fig.~\ref{fig-Tvt}) for several values of
the inelastic parameter $\gam$.  As the collisions become more
elastic, the decay rate approaches the theoretically predicted value
of $-2$ for a homogeneous cooling state.  Both axes are logarithmic,
and the line is a power-law fit, with a slope of 0.3.}
\end{figure}
\begin{figure}
\begin{center}\includegraphics{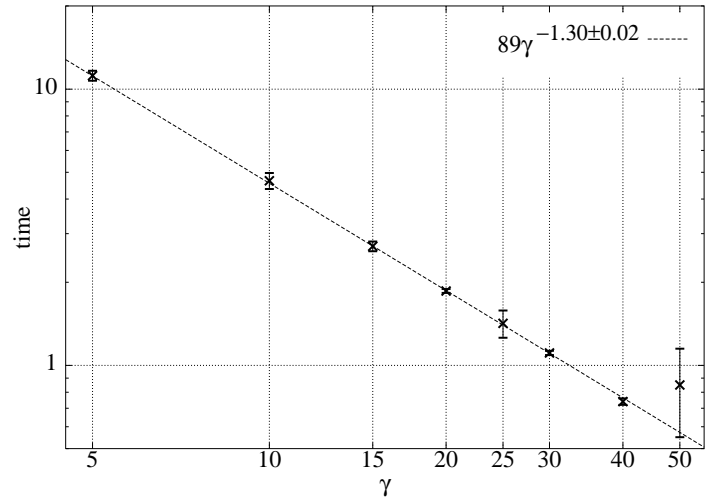}\end{center}
\caption{\label{fig-onset}The time at which the decay rate of the
temperature reaches maximum, as a function of $\gam$.  The errors are
rounding errors due to the finite sample rate.}
\end{figure}

Figure~\ref{fig-Tvt} shows the decay of the average temperature as a
function of time in our simulation, for three different values of the
inelasticity parameter $\gam$.  The initial decay approximately
follows the predicted $-2$ exponent, while for later times the
temperature decays at a slower rate as the instabilities agitate the
system\cite{LH99, NBC02}.  In the limit of low inelasticity, the
maximal rate of decay more closely approaches Haff's predicted
inverse-square behavior (Fig.~\ref{fig-approachHaff}).  (Note,
however, that the temperature will not decay at all in a completely
elastic system.)  In more inelastic systems, the hydrodynamic
instabilities kick in sooner and compromise the homogeneity of the
system.  One could characterize the time it takes for the
instabilities to emerge by the time it takes for the temperature to
reach its fastest decay rate.  Figure~\ref{fig-onset} shows that this
onset time decreases with respect to the inelasticity parameter as a
power-law with an exponent of $-4/3$.

\begin{figure}
\begin{center}\includegraphics{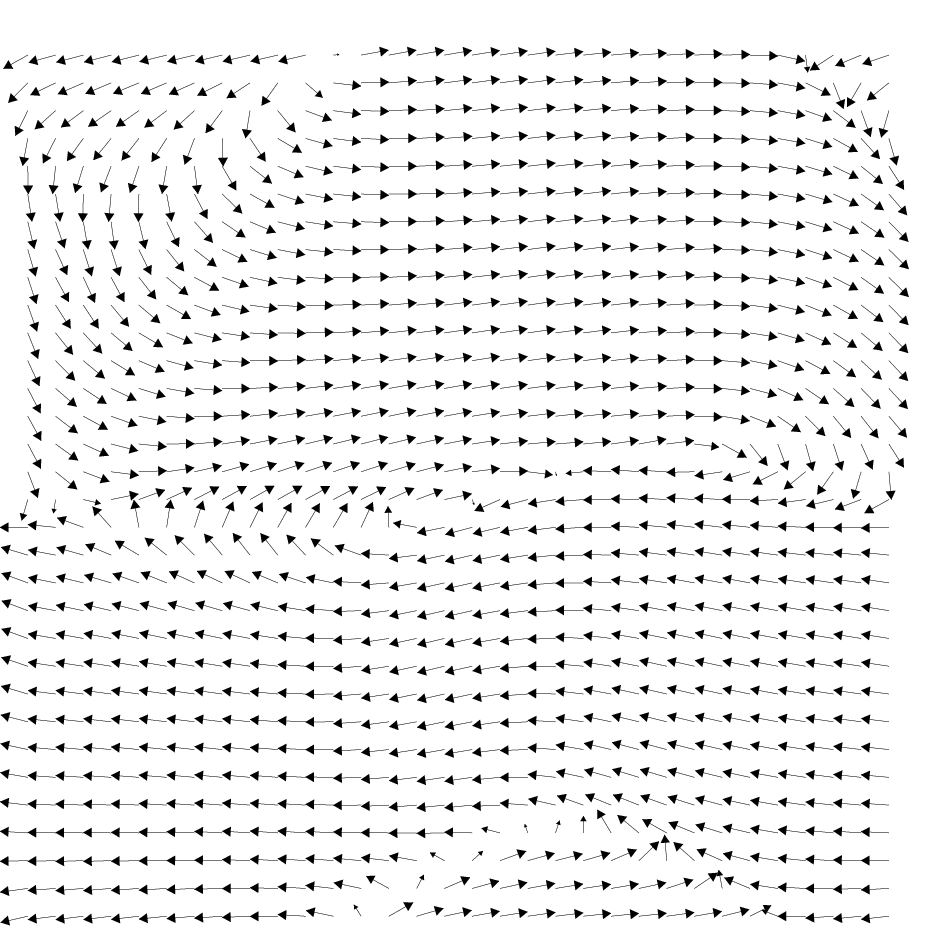}\end{center}
\caption{\label{fig-shear}A flow diagram for $\gam=50$ at $t=450$.
Each arrow represents the average velocity of four lattice sites, to
improve readability.}
\end{figure}

The first instability that is predicted to dominate the homogeneous
solution is a hydrodynamic shearing mode: two bands of material moving
in opposite directions.  This has been seen in several molecular
dynamic simulations\cite{DB97, McY96}, Fig.~\ref{fig-shear} shows how it has
developed in our system as two horizontal shear bands.
\begin{figure}
\begin{center}\includegraphics{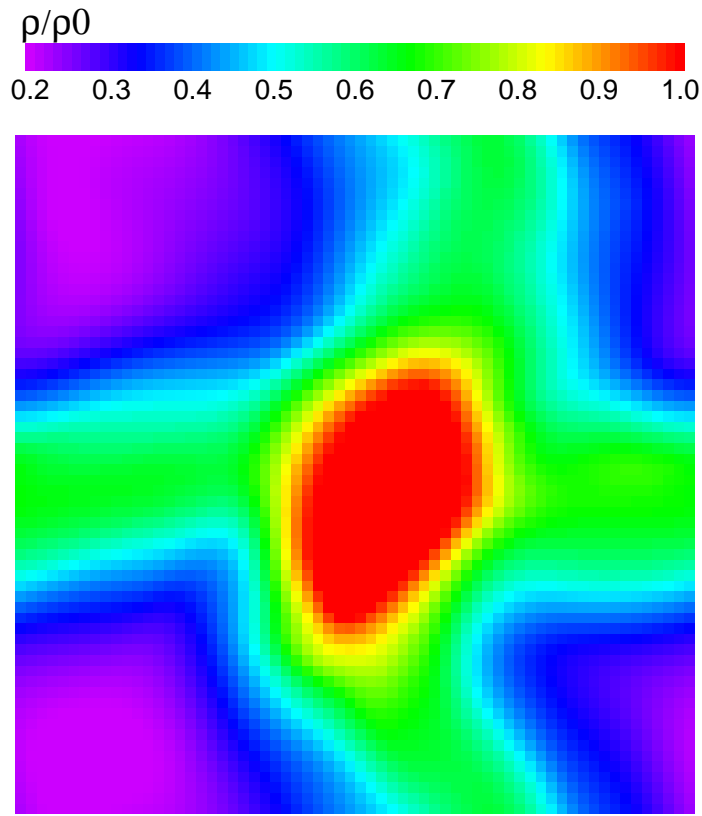}\end{center}
\caption{\label{fig-small}The density distribution for the $\gam=50$
system at time $t=100$.}
\end{figure}
\begin{figure}
\begin{center}\includegraphics{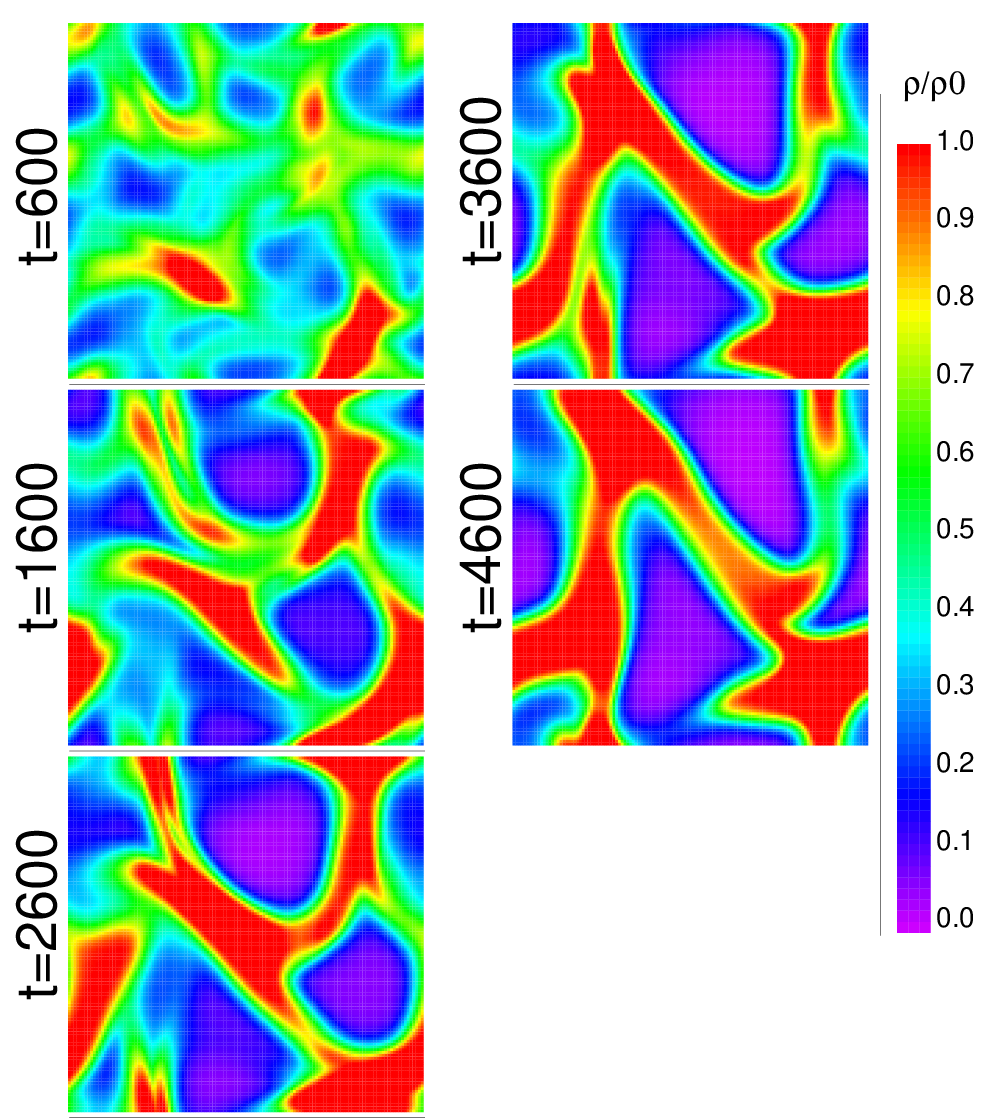}\end{center}
\caption{\label{fig-large}The density distribution for a large
($128\times128$) system at several times during its evolution.}
\end{figure}
Figure~\ref{fig-small} shows that our system develops a clustering
instability as well.  Note that the single cluster takes on a compact
shape, which is surprising given that there is no surface tension in
our model.  If clusters are supposed to grow by accretion, then one
would expect a ramified structure.  This puzzle becomes clearer as we
consider a larger system after further evolution.
Figure~\ref{fig-large} shows that clusters first form in these compact
shapes, but as time goes on they reach out to their neighbors,
stretching into the more stringlike forms seen in
simulation\cite{goldhirsch}.  In hindsight, we are able to see this
behavior in the smaller system as well, where the single cluster
interacts with itself through the periodic boundaries.

\begin{figure}[!htp]\begin{center}\includegraphics{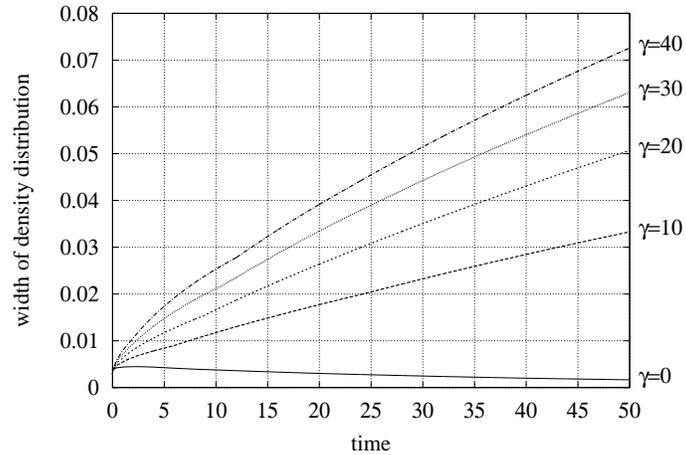}\end{center}
\caption{\label{fig-elastic}The width of the density distributions
(i.e. $\rho_{max}-\rho_{min}$) for several values of $\gamma$,
including the elastic case $\gamma=0$. Notice that the density
distribution is collapsing to a delta function in the elastic case,
approaching complete homogeneity, while the inelastic systems show 
broadening density distributions due to the clustering instability.}
\end{figure}

Finally, to demonstrate that this clustering instability is the result
of the inelastic parameter, we compare the width of the density
distribution for inelastic systems with the distribution for the
elastic case $\gam=0$ (Fig.\ref{fig-elastic}).  In the absence of
inelasticity, the density distribution collapses to a delta function,
indicating complete homogeneity.

Our results show that the shearing and clustering instabilities,
identified by Goldhirsch and Zanetti using a simplified version of the
above equations, exist in the complete nonlinear granular hydrodynamic
equations \ref{eq-eqns}.  Haff's cooling law is obeyed in the limit of
small inelasticity, but in general the instabilities become relevant
before the system has a chance to completely homogenize.  The
power-law dependence of the onset time for these instabilities on the
inelasticity, and the $-4/3$ exponent in particular, are interesting;
we have not found any reference to these in the literature.  Also
interesting is the way in which these clusters develop from compact
structures into networks which span the system.  It is not clear
whether an individual cluster begins to stretch out because of the
proximity of its neighbors, or because of effects due to its
increasing surface size.  One can imagine that the behavior of this
system could change as we alter the total amount of mass in the
system.

We thank Professor Todd DuPont for his assistance in constructing our
simulations, and Professors Heinrich Jaeger, Sidney Nagel, and Thomas
Witten for helpful conversations.  This work was supported by the
Materials Research Science and Engineering Center through Grant
No. NSF DMR 9808595.


\begin{thebibliography}{99}
\bibitem{temperature}The temperature $T$ here is not thermal, but
actually a measure of the local energy density ${\cal E}$ that is not
accounted for by the local flow velocity: $\rho T\equiv {\cal
E}-{1\over2}\rho u^2$.  Apart from this equation, it is not necessary
to assume that this quantity $T$ has any properties normally
associated with temperature.
\bibitem{fn-turbulence}The question is similar to whether all of the
physics of turbulence is captured by the Navier-Stokes equations, or
whether one must, at some point, resort to a more microscopic theory.
\bibitem{fn-method}Some details about our methods:
(1) Our Runge-Kutta method includes an adaptive time step; it,
however, is rarely activated since we use a default time step that is
suitably small.
(2) We offset the velocities from the other fields on the lattice
such that the densities and temperatures may be considered to lie on
the faces of the lattice while the horizontal (vertical) velocities
lie on the vertical (horizontal) edges.
(3) We have had great numerical difficulty with regions of very low
density: such regions have a tendency to drift below zero density or
generate large velocities in an unnatural fashion.  We use an upwind
technique to evaluate the equation $\dot\rho=-\nabla(\rho u)$, to
prevent the density from becoming negative.  In addition, we prevent
mass from flowing out of any site with a density below some minimum
value ($\rho_{\min}=0.001$).
\bibitem{fn-units}Our dimensionless density $\rh$ should not be
confused with the dimensionless packing fraction.  For numerical
convenience, our units of density have been scaled so that the random
close-packed density corresponds to $\rh_0=0.2$.  We will refer often
to the ratio $\rh/\rh_0$ to avoid confusion.
\end{thebibliography}
\end{document}